\documentclass[preprint]{nature}%
\usepackage{graphicx}
\usepackage[usenames,dvipsnames]{color}
\usepackage{siunitx}
\usepackage[colorlinks=true,citecolor=blue,linkcolor=blue,urlcolor=blue]%
{hyperref}
\usepackage[a4paper]{geometry}
\usepackage{fixltx2e}
\usepackage{upgreek,xspace}
\usepackage{amsmath}
\usepackage{amssymb}
\usepackage{amsthm}
\usepackage{mathrsfs}
\usepackage{amsbsy}
\usepackage{amstext}
\usepackage{xcolor}
\usepackage{float}
\usepackage{multirow}
\usepackage{wasysym}
\usepackage{mathtools}
\usepackage{subfigure}
\usepackage{amsfonts}
\usepackage{dcolumn}
\usepackage{float}
\usepackage{caption}
\usepackage[right]{lineno}
\usepackage{fancyhdr}
\usepackage{bbold,bm}
\usepackage[utf8]{inputenc}
\usepackage[T1]{fontenc}%
\providecommand{\U}[1]{\protect\rule{.1in}{.1in}}
\DeclareCaptionLabelSeparator{vline}{ $\boldsymbol{\vert}$ }
\usepackage{amssymb, amstext, amsmath}  % blackboard math symbols
\usepackage{physics}                    % easy quantum mechanics, some overlap with amsmath
\usepackage{bm}                         % bold math
\usepackage{nicefrac}                   % compact symbols for 1/2, etc.
\usepackage{tabularx,booktabs}          % professional-quality tables
\usepackage{soul,xcolor}                % coloring and strikethrough fonts
\usepackage{verbatim}                   % block comments in text
\usepackage{float,placeins}             % precision placement of figures

\def\be{\begin{eqnarray}}  
\def\ee{\end{eqnarray}}

\begin{document}

\title{All-optical fluorescence blinking control in quantum dots with ultrafast mid-infrared pulses}

\author{Jiaojian Shi$^{1, \dagger}$, Weiwei Sun$^{1, \dagger}$, Hendrik Utzat$^{1,2, \dagger}$, Ardavan Farahvash$^{1}$, Frank Y. Gao$^{1}$, Zhuquan Zhang$^{1}$, Ulugbek Barotov$^{1}$, Adam P. Willard$^{1}$, Keith A. Nelson$^{1,*}$, Moungi G. Bawendi$^{1,*}$}

\maketitle
\begin{affiliations}
	
\item Department of Chemistry, Massachusetts Institute of Technology, 02139 Cambridge, Massachusetts, United States
\item Present Address: Department of Materials Science and Engineering, Stanford University, 94305 Stanford, California, United States
\\
	$^{\dagger}$These authors contributed equally to this work\\
	$^{*}$Email: mgb@mit.edu kanelson@mit.edu
\newpage
	\vspace{1cm}
\end{affiliations}
\normalsize
\begin{abstract}
Photoluminescence (PL) intermittency is a ubiquitous phenomenon detrimentally reducing the temporal emission intensity stability of single colloidal quantum dots (CQDs) and the emission quantum yield of their ensembles. Despite efforts for blinking reduction via chemical engineering of the QD architecture and its environment, blinking still poses barriers to the application of QDs, particularly in single-particle tracking in biology or in single-photon sources. Here, we demonstrate the first deterministic all-optical suppression of quantum dot blinking using a compound technique of visible and mid-infrared (MIR) excitation. We show that moderate-field ultrafast MIR pulses (5.5 $\mu$m, 150 fs) can switch the emission from a charged, low quantum yield 'grey' trion state to the 'bright' exciton state in CdSe/CdS core-shell quantum dots resulting in a significant reduction of the QD intensity flicker. Quantum-tunneling simulations suggest that the MIR fields remove the excess charge from trions with reduced emission quantum yield to restore higher brightness exciton emission. Our approach can be integrated with existing single-particle tracking or super-resolution microscopy techniques without any modification to the sample and translates to other emitters presenting charging-induced PL intermittencies, such as single-photon emissive defects in diamond and two-dimensional materials.
\end{abstract}
\newpage

\newpage
%%%%%%%%%%%%%%%%%%%%%%%%
Colloidal quantum dots (CQDs) have now made significant commercial inroads in diverse optoelectronic applications as well as biological imaging due to their unique electronic structures, size-tunable emission, high photoluminescence (PL) quantum yield, high photo-stability, and facile chemical synthesis.\cite{Murray2000} However, despite two decades of research, stochastic PL intermittency, also known as 'blinking', still reduces the temporal emission stability, particularly under high excitation-flux conditions in single-emitter experiment, posing a significant barrier to the wider adoption of QDs in single-emitter applications. In single-photon sources, blinking reduces the ability of QDs to produce single-photons on demand and often coincides with detrimental spectral jumping\cite{Nirmal1996,Frantsuzov2008,Efros2016,Sauter1986,Dickson1997}, two processes that need to be eliminated to qualify QDs as building blocks of single-photon sources for quantum cryptography and computing.\cite{Kimble2008,Michler2000} In real-time single-particle tracking in biological systems\cite{Michalet2005,Dahan2003}, the timescale of blinking coincides with the particle diffusion reducing the tracking ability.

Despite the lack of a unified blinking theory\cite{Frantsuzov2008}, it is now understood that blinking in QDs occurs primarily via two pathways i) surface trap-mediated non-radiative recombination and ii) charging-induced Auger recombination.\cite{Galland2011} Synthetic efforts have been made towards intrinsically non-blinking quantum dots targeting both of these mechanism primarily through passivating surface trap-states and reducing the Auger recombination rate of charged states through smoothing of the potential barrier between core- and shell-layers of heteroepitaxial QDs.\cite{Chen2008,Mahler2008,Garcia2011,Ji2015,Chen2013,Nasilovwski2015} Extrinsic blinking control has been comparatively less explored. Adding anti-blinking agents suppresses blinking in QDs by passivating the QD surface.\cite{Hohng2004,Fomenko2008,Thomas2018} Other approaches such as electrostatic gating have also been reported to tune the Fermi level of the QD and block the relaxation pathways through the surface, thereby suppressing non-radiative decay during OFF periods.\cite{Jha2010,Galland2011} These methods, however, require the QD to be either deposited on a special substrate or immersed in a non-native environment, adding significant complexity to single-QD devices and precluding blinking control in biological environments. To date, no non-invasive active suppression of blinking has yet been demonstrated.

Here, we demonstrate the first deterministic all-optical method for active reduction of QD blinking with ultrafast MIR pulses. We build on advances in ultrafast electric-field pulse technologies\cite{Hebling2008,Manzoni2010} and show that off-resonant laser pulses at mid-infrared (MIR) frequencies provide sufficient strengths to overcome potential barriers in QDs while avoiding dielectric breakdown that often occurs upon applications of static electric fields. By applying MIR pulses concurrently with optical excitation of single QDs, we demonstrate that MIR pulses with an appropriate field strength can remove the excess electron of charging-induced blinking OFF states in single QDs. Therefore, MIR pulses have the potential to transiently discharge QDs in the blinking OFF states at an ultrafast speed without perturbing the equilibrium emissivity or introducing inter-band excitations.

The experimental setup is shown in Fig.1a. Our experiments investigate if the MIR pulse can switch OFF states to ON states by discharging the trion state (as shown in Fig. 1b) through the measurement of blinking statistics, emission lifetimes, and measuring emission spectra and emission intensities. A 1-kHz MIR pulse train at 5.5 $\mu$m is used in our experiments (Fig. 1c), which is far below the multiphoton absorption regime and away from phonon absorption frequencies of CdSe/CdS QDs.\cite{Nor2010} We tuned the MIR exposure by modulating a mechanical shutter and quantify the exposure by the number of pulses in a burst of MIR. Under 405 nm continuous-wave (CW) laser excitation, emission from single core-shell CdSe/CdS QDs with an 8-monolayer (ML) shell thickness, deposited on a glass coverslip, was confirmed by the antibunching dip ($\tau = 0$) from a second-order photon intensity correlation measurement, as shown in Fig. 1d.

We show the effect of concurrent MIR and optical excitation in Fig. 2. We record the PL intensity traces of single QDs under 405 nm CW laser excitation. Under no MIR field, we show a representative single QD PL blinking trace (Fig. 2a) and its histogram of PL intensity distributions (Fig. 2b). The histogram in Fig. 2b shows a bimodal PL intensity distribution, corresponding to blinking ON and OFF states. With MIR excitation at a suitable field strength (here $F/F_{max} = 0.6$, $F_{max}\sim10$ MV/cm), the blinking behavior from the same QD changes as shown in Fig.2c-d, reflected by the significant decrease in the time that the dot spends in the OFF state in the blinking trace as well as by the shift in the intensity histogram from bimodal to a largely unimodal ON state.

The effective conversion from OFF to ON states is also reflected by an enhanced PL intensity in ensemble QDs. By imaging and selecting isolated single QDs within the field of view, we can study the dynamics and responses of statistically-averaged PL intensity (counts summed over all the dots) when MIR excitations are turned on and off. Here, MIR excitations are periodically exposed on the sample for 200 ms (a burst of 200 MIR pulses) every 5 seconds, resulting in PL intensity spikes every 5 s, as shown in Fig. 2e. No degradation in either equilibrium PL counts without MIR or enhanced PL counts with MIR is observed, suggesting the reversible nature of the PL enhancement. The enhancement of PL intensity only appears when a suitable MIR field strength is applied. With stronger MIR pulses ($F/F_{max}>0.7$) , we observe a suppression of the overall PL intensity. In Fig. 2f, we show a drop in the PL intensity under an MIR exposure of 10 ms with a burst of 10 MIR pulses every two seconds at $F/F_{max}=0.9$, as shown in Fig. 2f.

Figure 2g illustrates the normalized PL intensity change as a function of CW-laser optical power. The MIR-induced PL intensity change is summed over multiple ($\sim$ 100) isolated dots and normalized to their equilibrium PL intensity. The PL percentage change increases as the optical power increases and saturates at $\sim$ 30 W/cm$^2$. The extrapolated zero PL change at zero optical power indicates that the MIR itself does not produce any luminescence and only enhances PL when the dots are excited optically. The positive correlation between the PL percentage change and optical power suggests that the PL enhancement is most likely due to removal of accumulative events such as excess photo-ionized charges induced blinking.\cite{Cherniavskaya2003,Galland2011} The threshold behavior in the MIR field dependence of the PL percentage change (Fig. 2h) further suggests that the MIR field is driving ionization processes and, as a result, removes excess charges inside the QDs. A crossover behavior is observed at field strengths around $F/F_{max}=0.7$ with an optical power of 20 W/cm$^2$ and a burst of 80 MIR pulses every 1 s, which is consistent with the observation of PL quenching under excessively high MIR fields shown in Fig. 2f. Figure 2i displays the MIR exposure dependence of the PL percentage change with a fixed optical power at 20 W/cm$^2$ and a field strength at $F/F_{max}=0.6$. The PL percentage change increases with increasing MIR pulse exposure. The PL percentage change saturates at a burst of around 100 MIR pulses every 1 s at the field strength of $F/F_{max}=0.6$. Further increasing MIR exposure results in PL intensity degradation, which can be explained by excess charge generation with excessive pulses.

PL lifetime changes further validate that MIR fields remove excess charges inside the QDs. When the MIR is off, a single QD randomly switches between ON and OFF states. The strong correlation between the PL intensity and lifetime indicates that blinking is caused by charging and discharging as expected for core-shell CdSe/CdS QDs\cite{Galland2011,Javaux2013} as is seen in Fig. 3a through a fluorescence lifetime–intensity distribution (FLID) plot. With MIR excitation, the blinking OFF states disappear, and only the ON states are present with a long lifetime and high PL intensity, as shown in Fig. 3b. Another representation is provided in the Supplementary Information by interrogating all the photon arrival times relative to the excitation trigger. As shown in Fig. S8a, without MIR fields, the PL lifetime of the single QD is composed of fast and slow exponential decays components, which we assign to trion and exciton decay, respectively. With MIR excitations, the fast trion decay is completely suppressed, and only the slow exciton relaxation is left, implying that MIR fields efficiently remove the excess charge.

Figure 3c and d show that blinking statistics are strongly modified by MIR excitation. By setting a threshold in a blinking trace separating ON and OFF states and counting their probability densities, a power-law distribution of ON- and OFF-times is obtained in Fig. 3d and c, respectively. As a manifestation of self-similarity behavior that is seen in the blinking of many types of fluorophores\cite{Frantsuzov2008}, the power-law relationship is highly robust to external perturbations.\cite{Bharadwaj2011,Kuno2000,Frantsuzov2008} A previous study by Hasham \textit{et al.}\cite{Wilson2020} reported that a sub-bandgap CW laser can change the QD blinking power-law statistics by depleting excited states, resulting in more blinky QDs. In our experiments, we find that the MIR field depletes the OFF states and alters the OFF-time power-law statistics. Since the MIR irradiates the sample with an 80 ms burst of 80 pulses every 1 s, converting OFF states to ON states, OFF events longer than 1 s are strongly suppressed by over an order of magnitude, as shown in Fig. 3c. Probability cut-off times also emerge at the harmonics of the one-second excitation period. Besides the suppression of long OFF events, the MIR field also creates two short OFF events by splitting longer OFF events. As a result, we observe spikes in the probability density to the left side of the cut-off times at 1 s and at its harmonics. For ON-state statistics, more ON events longer than 1 s emerge.

A change in the PL spectrum averaged over multiple single dots under the MIR fields is observed. Figure 3e shows that the MIR field induces a spectral blueshift. We subtracted the MIR-off PL spectrum from the MIR-on spectrum for better visualization, as shown in Fig. 3f. The spectral blueshift manifests as the first derivative feature in the differential plot, and a second derivative component, i.e., spectral narrowing, can be deduced as well. We further obtain the change in the PL spectrum as a function of MIR field strengths in Fig. 3g, which exhibits a crossover behavior. At the moderate-field regime ($F/F_{max}=0.2-0.6$), the MIR field blueshifts the PL spectrum, restoring excitonic emission.\cite{Empedocles1999,Patton2003,Bracker2005} At the high field regime ($F/F_{max}>0.7$), the spectrum redshifts, indicating MIR fields not only remove those excess electrons in the trion state but also introduce additional charges to the QD by ionizing the exciton state.

We further examine the MIR responses on QDs with different shell thicknesses, including 10 and 14 ML QDs. We first characterize ensemble equilibrium PL. We observed a suppressed blinking behavior in giant-shell (14 ML) QDs consistent with the literature.\cite{Chen2008} Biexciton quantum yield (BXQY) measurements were done for 8, 10, and 14 ML QDs. The BXQY can be obtained from single-dot photon-correlation measurements but this can suffer from large heterogeneity\cite{Zhao2012,Park2011,Park2014} due to the high sensitivity of competing Auger non-radiative pathways to defects and local environments.\cite{Nair2011} Instead a solution-phase sample-averaged BXQY measurement was done to enable us to extract ensemble-averaged single-dot BXQY.\cite{Beyler2014} Figure 4b shows that the BXQY for 8 ML QDs is $5.4\pm1.0$\%. For 10 ML, the BXQY is $10.7\pm1.2$\% and it is $21.3\pm1.0$\% for 14 ML (Fig. S1). The higher BXQY in thicker-shell QDs suggests that the non-radiative pathways are of lesser importance in thicker shell QDs. In the ensemble PL lifetime measurement of QDs solution as shown in Fig. 4a, 14 ML QDs exhibit an extraordinarily long emission tail over tens of microseconds, while 8 ML QDs show a much shorter exponential decay. At longer decay times, the emission from 14 ML QDs deviates from an exponential and exhibits a power-law decay behavior (the fit is shown in Fig. S7), suggesting a power-law trap-detrap assisted delayed emission.\cite{Rabouw2015,Rabouw2020, Jones2009,Sher2008} In 14 ML QDs, the delayed emission contributes a significant portion (around 30\%) of the total emission. These results suggest that trion-mediated blinking is less important in 14 ML than in 8 ML QDs, and consequently, we expect the MIR response in 14 ML QDs to be different than that in 8 ML QDs. 

We now discuss the effect of shell-thickness on the MIR response. Figure 4c shows the dependence of MIR-induced PL intensity changes of 10 and 14 ML QDs under differing experimental conditions. 10 ML QDs exhibit qualitatively similar behavior with 8 ML QDs. As for the optical power dependence, the response of 10 ML QDs peaks at about three times lower optical power than 8 ML QDs, which is due to the difference in absorption cross-section for different shell-thicknesses (the absorption cross-section calibration is included in Fig. S4). The trion and exciton ionization thresholds in 10 ML QDs are slightly higher than in 8 ML QDs, likely because the tunneling barrier across the shell is wider in 10 ML QDs. 14 ML QDs, as mentioned above, exhibit quite distinct MIR responses than thinner shell QDs. At all field strengths, for 14 ML QDs, the MIR blueshifts the PL spectrum (Fig. 4d) and quenches PL intensity (Fig. 4c), contrary to the PL enhancement along with blueshift at moderate field strength in 8 ML QDs. As a result, we observe a reverse effect of PL dynamics change in 14 ML QDs than in 8 ML QDs. The blinking OFF-time distribution is almost unaffected in 14 ML QDs, while the ON-time distribution is altered at 1 second and at its harmonics, contrary to the nearly unaffected ON-time distribution and altered OFF-time in 8 ML QDs. Additional data can be found in the Supplementary Information (Fig. S5 and S6).

To explain the blinking suppression in 8 and 10 ML QDs, we constructed a one-dimensional model of the QD confined excess electron and use numerical simulations to calculate the response to the applied time-dependent MIR field. The core-shell structure of the QD is described by a stepwise potential with a central low energy well, representing the core, and a higher energy plateau, representing the shell, as illustrated in Fig. 5a. Trap states are modeled as an additional stepwise potential well residing at the periphery of the shell. Tunneling probabilities from the core to the trap regions are calculated over the course of an ultrafast MIR pulse by numerically integrating the time-dependent Schr\"{o}dinger equation under the dipole approximation (see Methods). Using a physically motivated selection of parameters, the model reproduces the observed field strength threshold behavior in thin shell QDs, as shown in Fig. 5b. The lower threshold can be understood through the interplay between the length of the pulse ($\tau_{pulse}$) and the spatial overlap between the core-localized and trap-localized electron eigenstates. The theoretical treatment at the high-field regime is more elusive as it requires explicit differentiation between trion and exciton states. In principle, such treatment requires a many-body model which explicitly accounts for interactions between electrons and holes \cite{nandan_wavefunction_2019}, however here we adopt a simpler approach by adding the exciton binding energy (estimated to be around 0.2 eV\cite{nandan_wavefunction_2019,elward_effect_2013}) to the energy difference between the core and the shell. The larger energy barrier reduces the spatial overlap between core and trap eigenstates, shifting the field threshold for tunneling to the trap to higher values. By subtracting trap tunneling probabilities with and without the additional exciton binding energy, we obtain a curve that qualitatively reproduces the experimental field dependence over the entire range of field strengths for both the 8 ML (Fig. 5b) and 10 ML QDs (Fig. S13).

In 8 ML QDs, we conclude that MIR fields remove excess charges in QDs and consequently change PL intensity and spectrum, as visualized in Fig. 5c, supported by both the experimental observations and simulations. In 8 ML QDs, blinking is caused by trion-mediated Auger recombination, corroborated by our observation of a strong correlation between PL intensity and lifetime.\cite{Galland2011} With moderate-field MIR pulses, the trion is ionized and the excess charge is removed, restoring an exciton state. Consequently, PL intensity increases, and spectrum blueshifts. We attribute this blueshift to both the restoration of exciton emission, which always has higher energy than the trion emission, and the removal of quantum-confined Stark effect (QCSE). The excess charge in the OFF state can create an internal electric field and induce a QCSE, which distorts the electron and hole wavefunctions and consequently redshifts and broadens the emission spectrum.\cite{Empedocles1997} As a result, MIR fields shift the emission to higher energy and narrow the spectrum. At a high-field regime, MIR fields not only relocate the excessive charges but also ionize the restored excitons, activating additional charge generation and subsequent trion formation that leads to Auger non-radiative processes. The charge left inside the QD again creates a local electric field and stark shifts the exciton emission, leading to a decreased PL intensity and redshifted spectra.

In 14 ML QDs, we attribute the PL change under MIR fields to the removal of trapped excitons, as depicted in Fig. 5d. The long tail power-law decay lifetime shown in Fig. 4a suggests that a significant portion (30\%) of exciton emission comes from trap-detrap assisted delayed emission. In thick-shell QDs, electrons display transitory carrier separation, which has been reported to be more likely than in thin shells\cite{Bae2013,Rabouw2015} due to a larger number of traps per QD and more substantial delocalization of the electron wavefunction into the shell.\cite{Rabouw2015} Thus, the electron can be transiently separated from the hole wavefunction and then slowly returns to the recombination center due to the giant shell, followed by a radiative recombination. Those charge-separated electrons are localized outside the core and tend to be readily removed by MIR fields. With MIR excitations, the trapped electrons are depleted and can no longer return to the recombination center, supported by a faster PL lifetime (Fig. S8c), therefore removing the delayed emission and decreasing the overall PL intensity. The delayed emission is usually of lower energy,\cite{Rabouw2015} the removal of which results in a blueshifted spectrum, consistent with our observations in Fig. 4d.

In conclusion, we demonstrate that ultrafast MIR electric-field pulses can effectively remove excess charges responsible for the trion-mediated Auger recombination in blinking OFF states, thereby suppress the PL blinking and achieve near-unity quantum yield even at very high excitation flux. At high field strengths, MIR fields can further ionize excitons and cause additional charging. Experimental and simulation results support that MIR fields can manipulate carriers in QDs. Our all-optical approach enables almost non-blinking thin-shell QDs in a native environment, which is highly desirable for real-time single-molecule tracking of biological processes.\cite{Michalet2005} For example, fast clathrin-independent endocytosis\cite{Boucrot2015,Watanabe2013} involves transmembrane delivery and has been in need of a non-blinking fluorescence tag. The excitation wavelength can be readily extended to other spectral ranges due to the generality of the field-driven ionization mechanism, opening in vivo applications as one can choose a wavelength with minimal environmental absorption. The ultrafast nature of the pulse further renders cumulative heating effects negligible. Our experimental results should also motivate a hitherto-unexplored class of all-optical blinking control experiments with off-resonant field excitations in single emitters beyond CQDs, including nitrogen/silicon-vacancy color centers in diamond\cite{Doherty2013} and defects in two-dimensional transition metal dichalcogenides\cite{Srivastava2015,He2015,Koperski2015,Chakraborty2015}. The potential realization of single quantum emitters free from interruptions can pave the way to potential quantum computing and quantum cryptography applications.

\newpage

\section*{Methods}

\noindent \textit{Single QD sample preparation}\\
To synthesize CdSe/CdS quantum dots, an established synthetic protocol from the literature was used.\cite{Carbone2007} To synthesize CdSe core quantum dots, 60 mg CdO, 280 mg octadecylphosphonic acid and 3 g trioctylphosphine oxide were combined in a 50 mL round bottom flask. The resulting reaction mixture was put under vacuum and heated to 150 $^{\circ}$C to remove volatile substances. After 1 hour, the mixture was heated further to 320 $^{\circ}$C under nitrogen flow to form a clear colorless solution, and 1.0 mL trioctylphosphine was added dropwise. The temperature was increased to 380 $^{\circ}$C and the heating mantle was removed. 0.5 mL Se/trioctylphosphine (60 mg Se in 0.5 mL trioctylphosphine) was injected rapidly and the reaction mixture was cooled down to room temperature with air. The crude reaction mixture was washed with acetone and redispersed in toluene. This synthetic protocol resulted in highly monodisperse core-only quantum dots with first exciton absorption at 487 nm. To overcoat CdSe core quantum dots with CdS shells, continuous injection synthesis was used.\cite{Chen2013} In a 100 mL round bottom flask, 100 nmol CdSe core QDs in toluene, 3 mL octadecene (ODE), 3 mL oleylamine and 3 mL oleic acid were mixed. The resulting mixture was degassed for 20 minutes at room temperature and for 40 minutes at 100 $^{\circ}$C to remove volatile substances. Afterwards, the mixture was put under nitrogen flow, and the temperature was increased. When the temperature reached 200 $^{\circ}$C, 0.08 M cadmium oleate-ODE and octanethiol-ODE (1.2 equivalents) were added at 2.5 mL/hour to form 7 monolayers of CdS shell. After the completion of shell precursor injection, the reaction mixture was annealed at 310 $^{\circ}$C for 15 minutes. Quantum dots were washed with acetone and redispersed in hexane three times. Dense colloidal QDs were diluted by a factor around $10^6$, and then were drop cast or spin coated on a cover glass for single-dot measurements.\\

\noindent \textit{Ultrafast mid-infrared pulse generation}\\
MIR pulses were generated in a 0.5 mm thick GaSe crystal by difference-frequency mixing of the signal and idler outputs of a high-energy optical parametric amplifier (OPA). The OPA was pumped with $\sim$ 35 fs pulses from a commercial Ti:Sapphire regenerative amplifier (800 nm central wavelength 800 nm, 12 mJ) and has an output power at signal and idler at 2.5 mJ and 2 mJ. The signal and idler beam sizes were shrinked by a reflective telescope by a factor of 1.5 to reach optimal difference-frequency generation (DFG) efficiency. The generated MIR pulses from the GaSe crystal were collected and expanded by a pair of 3-inch diameter parabolic mirrors before being tightly focused to the sample by a third parabola with a 3-inch diameter and 2-inch effective focal length. The MIR field strengths were controlled by a pair of wire-grid polarizers (Thorlabs WP25H-K).\\

\noindent \textit{Photoluminescence measurements}\\
For PL imaging, a continuous-wave laser beam (center wavelength 405 nm, Toptica iBeam smart) was used to illuminate the sample with an up to 150 W/cm$^2$ excitation fluence. The PL image was collected by a 100$\times$ objective with a numerical aperture (NA) of 0.7 and imaged on an electron-multiplying charge-coupled device (EMCCD, from Andor iXon Ultra). We used a home-built confocal epifluorescence microscope to select individual QDs for single dot PL lifetime and spectral measurements. For PL spectrum characterization, successive measurements of the spectra were recorded in a spectrometer (Andor Shamrock spectrometer and Andor Newton 920). For PL lifetime measurements, a 420 ps resolution (instrumental response function) was achieved by exciting the sample with a pulsed laser source (center wavelength 405 nm, repetition rate 5 MHz, 100 ps pulse duration, Picoquant) at around 2 $\mu$J/cm$^2$ and detecting the emission with an avalanche photodiode (ID Quantique, id100-vis). The lifetime was obtained by tagging the arrival time of each PL photon relative to the trigger photon using a time-to-digital converter (ID Quantique, id800-TDC). The duration of MIR irradiation was controlled by a mechanical shutter. All the above measurements were performed in an ambient environment.\\

\noindent \textit{Solution biexciton quantum yield measurements}\\
Solution biexciton quantum yield measurements were completed by utilizing methods described by Beyler \textit{et al}.\cite{Beyler2014} Samples were diluted in hexane by a factor of around $10^4$. One drop of 2 mM Cadmium oleate solution in hexane was added to reduce particle aggregation, producing an average occupation in the focal volume between 6 and 8. The solution was wicked into a rectangular capillary (VitroCom, 0.100 $\times$ 0.200 mm i.d.) and sealed with capillary tube sealant to prevent solvent evaporation. Samples were excited with a pulsed laser source (532 nm, Picoquant) at a repetition rate of 1 MHz via a home-built confocal epifluorescence setup. Excitation was focused into the sample, and emission was collected using the same infinity-corrected water-immersion objective (Nikon, Plan Apo VC 60$\times$ WI, NA 1.2). A 535 nm long-pass filter (Chroma Technology Corp) was used to separate excitation and emission. Emission was further filtered spatially by a 1:1 telescoping 50 $\mu$m pinhole (Thorlabs) and spectrally by a 532 nm notch filter (Chroma Technology Corp) before being sent to three 50:50 non-polarizing beamsplitters (Thorlabs) as described by Shulenberger \textit{et al.},\cite{BXQYsetup} creating four equivalent intensity beams. Each beam was focused onto a single-photon counting detector (PerkinElmer, SPCM-AQR13) by a 10 cm achromatic lens (Thorlabs). Photon arrival times were recorded by a HydraHarp 400 (Picoquant) and analyzed using home-built software published at https://github.com/nanocluster/photons.\\

\noindent \textit{Quantum tunneling calculation}\\
Electron dynamics were calculated via a one-body, one-dimensional model with step functions between the core, shell, and trap. The potential energy for free-electrons in the conduction band was given by,
\begin{equation}
V(r) =
	\begin{cases}
	0 & 0 \leq |r| \leq r_c\\ 
	0.32 \text{eV} &  r_c \leq |r| \leq r_s\\
	0.07 \text{eV} &  r_s \leq |r| \leq r_t\\ 
	4.4 \text{eV} & |r| \geq r_t 
	\end{cases},
\end{equation}
where $r_c$, $r_s$, and $r_t$ are the outer edges of the core, shell, and trap respectively. These parameters are provided in Supplementary Information. For both the 8 ML and 10 ML QDs, the core-shell conduction band energy shift of 0.32 eV was taken from several sources.\cite{nandan_wavefunction_2019,steiner_determination_2008,sitt_multiexciton_2009,panfil_electronic_2019,raino_probing_2011} The core-trap offset of 0.07 eV $= 3k_B T$ was informed by the observation that electrons can stochastically tunnel between the core and trap at room temperature, suggesting that the energy difference should be on the order of $k_B T$.\cite{Jones2009,giansante_surface_2017} Finally, a vacuum ionization energy of 4.4 eV was applied for distances beyond $r_t$.\cite{mattoussi_electroluminescence_1998} In order to simulate electron tunneling in the exciton state, the core-shell energy shift was increased to 0.52 eV to roughly account for the exciton binding energy.\cite{nandan_wavefunction_2019,elward_effect_2013}

The lowest 70 eigenstates of this stationary potential were calculated using the matrix numerov method.\cite{pillai_matrix_2012} The ground state and first 3 dipole-allowed states are shown in Supplementary Information (Fig. S10). Time evolution under the MIR field was done by integrating the time-dependent Schr\"{o}dinger equation using the eigenstates of the stationary potential and a semiclassical representation for the laser field in the Coulomb gauge with the dipole approximation,
\begin{equation}
	\dot{c}_{k}(t) = -\frac{i}{\hbar }E_k c_k(t) - \frac{1}{\hbar} \sum_{l} F(t)  \frac{\omega_{kl}}{\omega } \mel{\psi_k}{ \hat{\mu} }{\psi_l} c_l(t).
\end{equation}
Here $c_k$ is the coefficient of the $k$th eigenstate, $\omega_{kl}$ is the frequency difference between the $k$ and $l$ states, and $\omega$ is the frequency of the field: $1818$ $\text{cm}^{-1} = 0.2254$ $\text{eV}$. A $\sin^2$ envelope was used to approximate the pulse waveform,
\begin{equation}
 	F(t) = |F| \sin^2 \left( \frac{\pi t}{\tau_{pulse}} \right) \sin(\omega t),
\end{equation}
where $\tau_{pulse}=150$ fs is the approximate length of the pulse. The simulations were run for a total of 300 fs, where the field was only turned on for the first 150 fs. All observables of interests were extracted by averaging over the second 150 fs.

\subsection{Data availability.}

The data that support the findings of this study are available from the corresponding author upon request.

\newpage

\section*{References}

\footnotesize
\providecommand{\noopsort}[1]{}\providecommand{\singleletter}[1]{#1}%

\newpage
\normalsize

\begin{addendum}
\item[Supplementary Information] Supplementary information is available in the online version of the paper.

\item[Acknowledgements] J.S., A.F., F.Y.G., Z.Z., U.B., A.P.W., K.A.N., and M.G.B. acknowledge support from the U.S. Army Research Lab (ARL) and the U.S. Army Research Office through the Institute for Soldier Nanotechnologies, under Cooperative Agreement number W911-NF-18-2-0048. W.S., H.U., and M.G.B. acknowledge support from the U.S. Department of Energy, Office of Basic Energy Sciences, Division of Materials Sciences and Engineering (award no. DE-FG02-07ER46454). J.S., F.Y.G., Z.Z., and K.A.N. acknowledge additional support from the Samsung Global Outreach Program.

\item[Author Contributions] 
J.S. and H.U. conceived of the study and initiated the experiments. J.S., F.Y.G., and Z.Z. conducted and refined the IR excitation measurements, developed the data acquisition and image processing software, and analyzed the experimental data. W.S. performed the solution-biexciton experiments and contributed to data analysis and interpretation. A.F. performed the quantum tunneling simulations under the supervision of A.P.W. U.B. synthesized CdSe/CdS quantum dots. J.S., W.S., and H.U. led the manuscript preparation. M.G.B. and K.A.N. supervised the project.

\item[Author Information] The authors declare no competing financial interests. Correspondence and requests for materials should be addressed to M.G.B. and K.A.N.
\newpage

\begin{center}
\includegraphics[width=\textwidth]{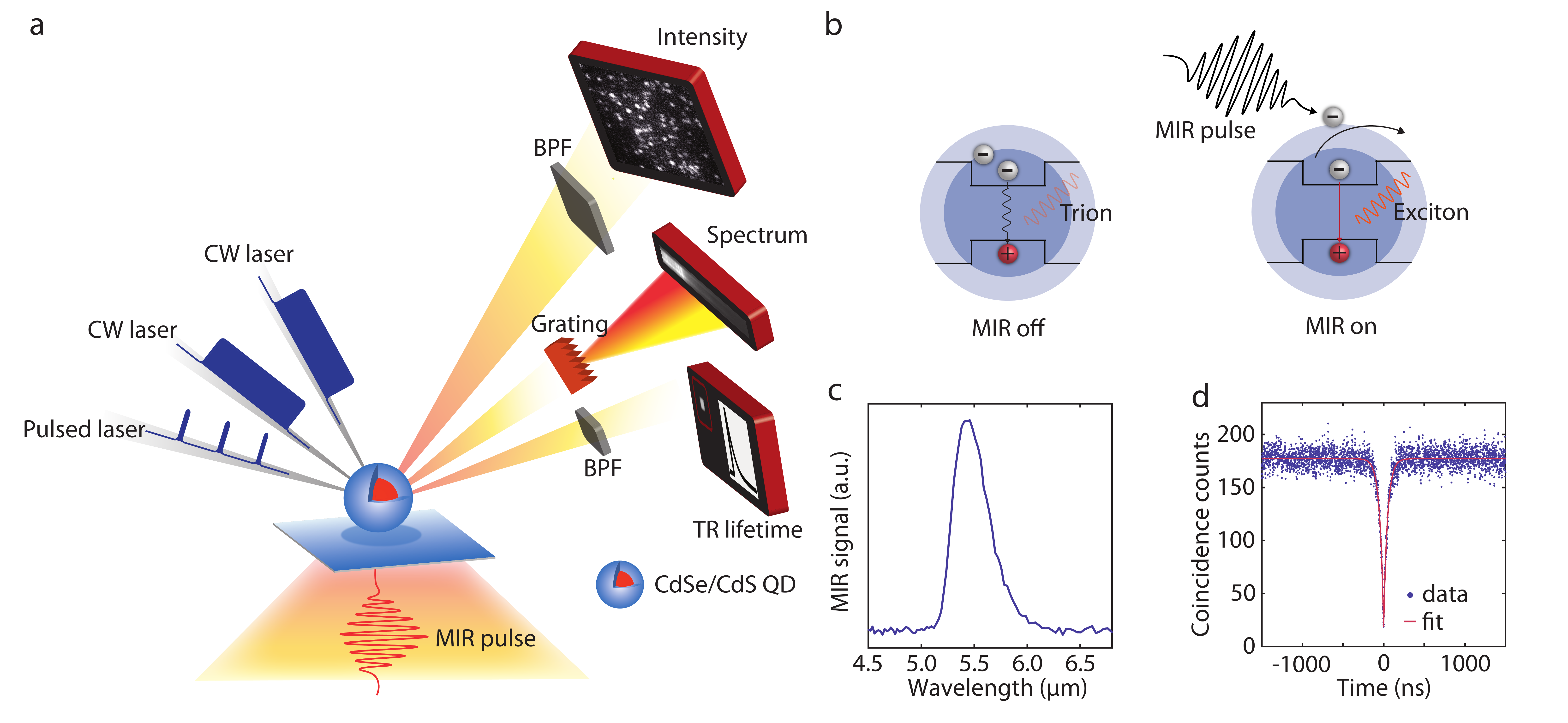}
\label{fig:Fig1}
\end{center}
\item[Fig. 1] \textbf{Experimental scheme and single dot verification.} \textbf{a,} Schematic illustration of a single CdSe/CdS quantum dot (QD) with mid-infrared pulse excitation and various PL probes, including PL intensity, PL spectrum, and (with pulsed rather than CW photoexcitation) time-resolved (TR) PL lifetime measurements. BPF: bandpass filter. \textbf{b,} In the conventional charging model for PL blinking, ON and OFF periods correspond to a neutral nanocrystal (exciton) and a charged nanocrystal (trion), respectively. During the OFF periods, ultrafast MIR fields can effectively remove the excess charge in trion and convert it to an exciton. \textbf{c,} The spectrum of MIR pulses centers at $\sim 5.5$ $\mu$m with a bandwidth $\sim 0.5$ $\mu$m. \textbf{d,} The second-order PL intensity correlation function measured for a single QD indicates that $g_2(0) = 0.1$.

\newpage
\begin{center}
\includegraphics[width=\textwidth]{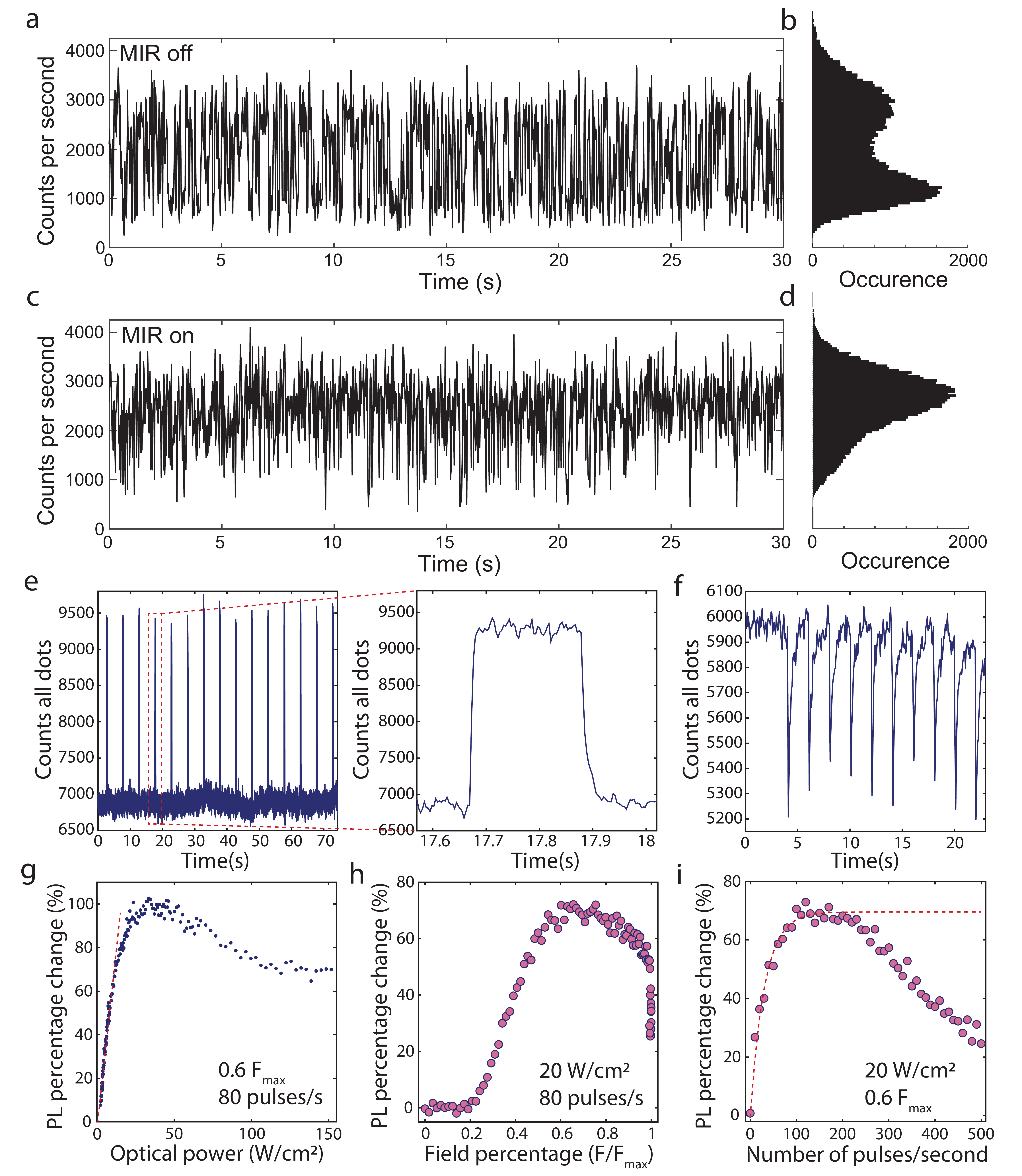}
\label{fig:Fig2}
\end{center}
\item[Fig. 2] \textbf{MIR blinking control in single QDs.} \textbf{a,} Representative PL blinking trace of a single QD without MIR excitations. \textbf{b,} The histogram indicates the distribution of intensities observed in the trace without MIR excitations. It shows bimodal distribution with ON and OFF events. \textbf{c,} Representative PL blinking trace of a single QD (the same QD of \textbf{a}) with MIR excitations. \textbf{d,} The histogram indicates the distribution of intensities observed in the trace with MIR excitations. It only has ON time fractions. \textbf{e-i,} MIR control of PL intensities averaged over multiple ($\sim$100) isolated QDs. \textbf{e,} PL counts averaged over all dots as a function of time. MIR pulses with a field strength of $F/F_{max}=0.4$ are turned on for 200 ms (a burst of 200 pulses) every 5 s, with a zoom-in view of PL enhancement in a burst of MIR shown at right. \textbf{f,} Averaged PL counts as a function of time with a field strength of $F/F_{max}=0.9$ and MIR exposure of 10 ms (a burst of 10 pulses) every 2 s. \textbf{g,} Optical power dependence of the normalized MIR-induced PL intensity change with a burst of 80 MIR pulses every 1 s at $F/F_{max}=0.6$. \textbf{h,} Field dependence of MIR-induced PL intensity percentage change at a 20 W/cm$^2$ optical power and MIR exposure (a burst of 80 MIR pulses) every 1 s. \textbf{i,} MIR exposure (number of pulses burst every 1 s) dependence of MIR-induced PL intensity percentage change at an optical power of 20 W/cm$^2$ and a MIR field of $F/F_{max}=0.6$.

\newpage
\begin{center}
\includegraphics[width=\textwidth]{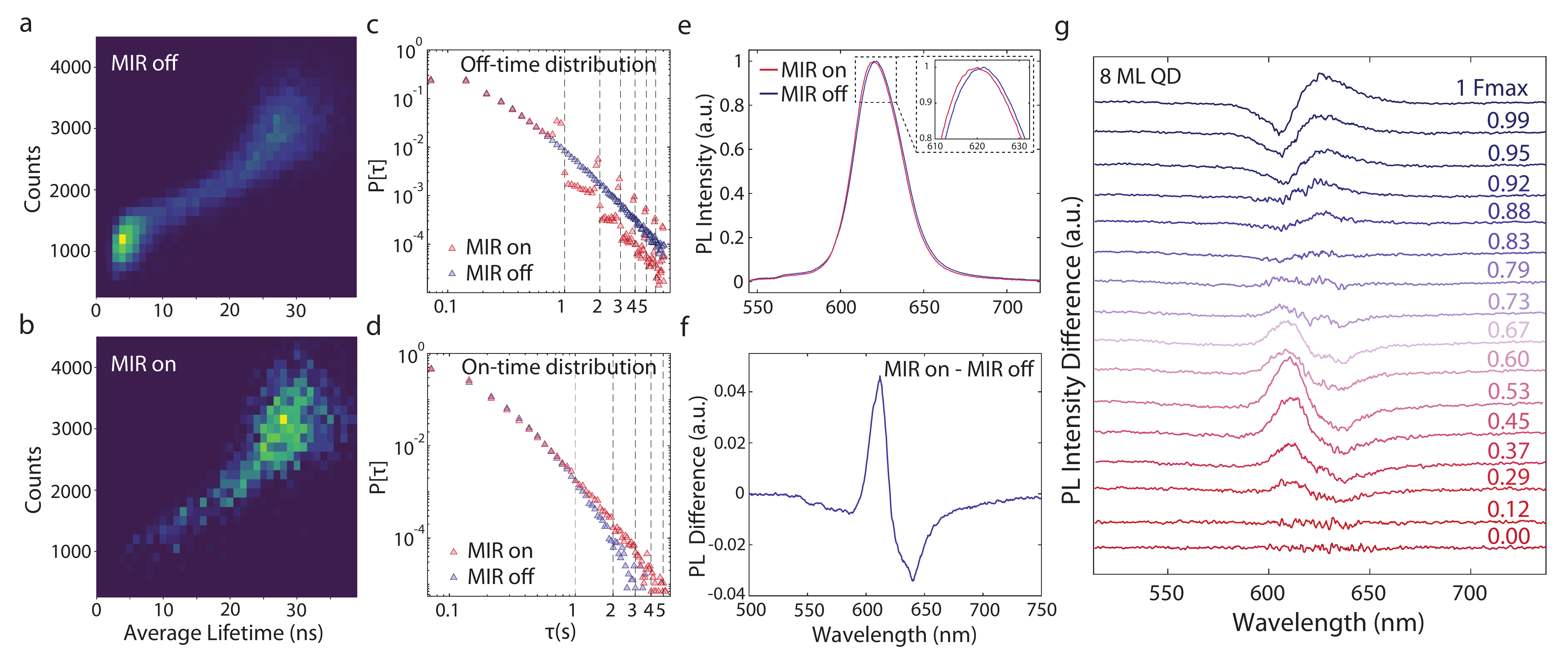}
\label{fig:Fig3}
\end{center}
\item[Fig. 3] \textbf{MIR fields alter PL lifetime, blinking statistics, and spectrum of single QDs.} \textbf{a-b,} Fluorescence lifetime–intensity distribution of a single QD with and without MIR excitations. Without MIR excitations, the single dot blinks between ON and OFF states, and the PL intensity strongly correlates with the lifetime. With MIR excitations, only the ON state with a long lifetime is present. \textbf{c,} OFF-time blinking statistics for isolated single QDs at equilibrium (blue) and during MIR excitations every one second and with a burst of 80 pulses in each period (red). Probabilities for OFF events longer than 1 s and its harmonics are strongly suppressed and converted to short blinking OFF events. \textbf{d,} ON-time blinking statistics for isolated QDs at equilibrium (blue) and during MIR excitations (red). Probabilities for ON events longer than $\sim$1 s are enhanced. \textbf{e,} PL spectrum without and with MIR excitations. The inset is the closeup of the spectral shift. \textbf{f,} Differential PL spectrum (MIR on - MIR off) shows a spectral blueshift and narrowing. \textbf{g,} Field dependence of MIR-induced spectral changes shows a crossover behavior.

\newpage
\begin{center}
	\includegraphics[width=\textwidth]{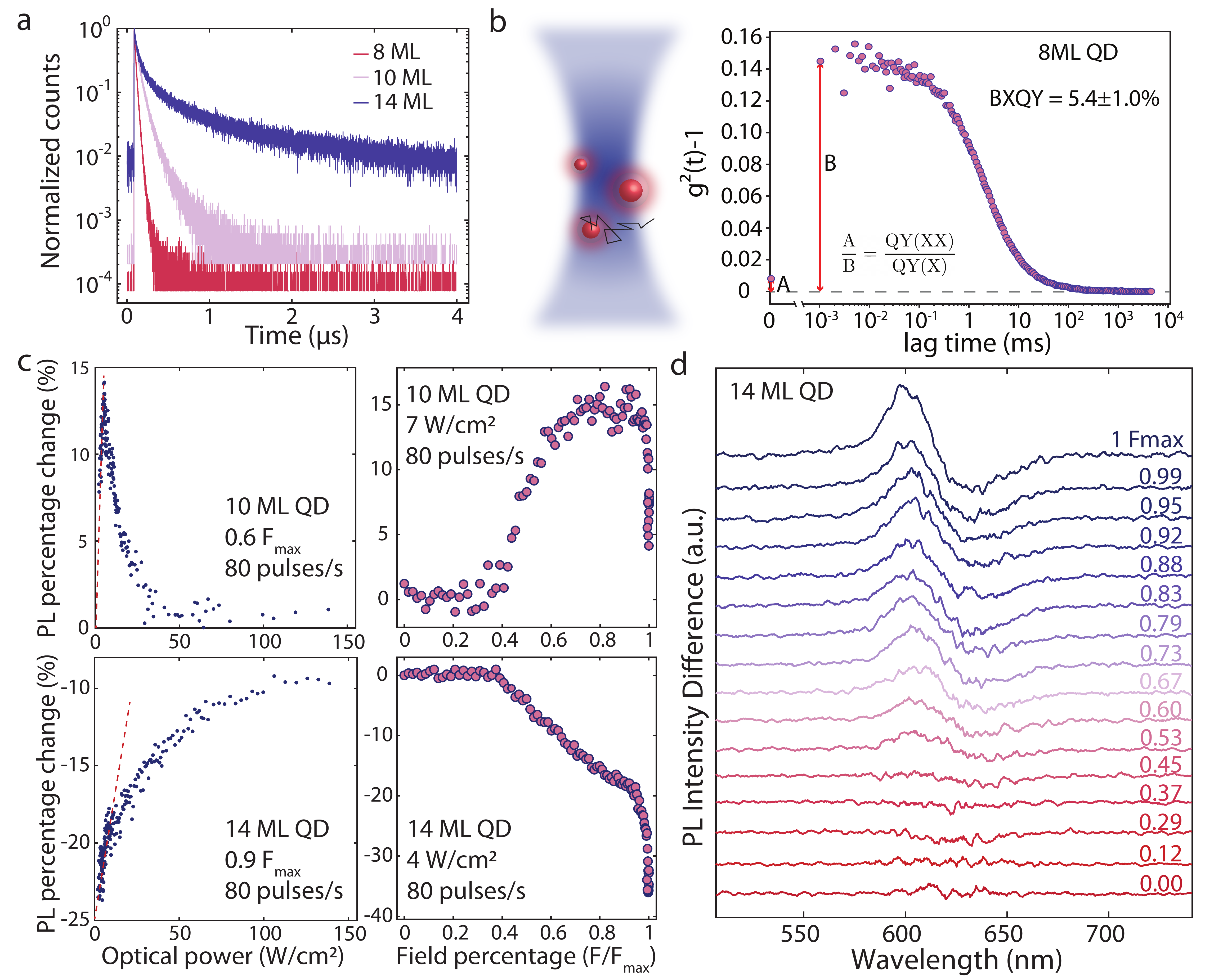}
	\label{fig:Fig4}
\end{center}
\item[Fig. 4] \textbf{MIR responses of QDs with different shell thicknesses.} \textbf{a,} PL lifetime of ensemble QDs with 8 (red), 10 (pink), and 14 (blue) monolayer (ML) shell thickness. 14 ML QD exhibits power-law delayed emission with a long tail in decay times. \textbf{b,} Solution biexciton quantum yield (BXQY) data of 8 ML QDs. BXQY for 8 ML QD is $5.4\pm1.0$\%. BXQY for 10 ML QDs is $10.7\pm1.2$\% and $21.3\pm1.0$\% for 14 ML QD (data in Fig. S1). \textbf{c,} Optical power and MIR field strength dependences in 10 and 14 ML QDs. \textbf{d,} Field dependence of the MIR-induced spectral shift in 14 ML QDs. MIR quenches PL in 14 ML QDs and introduces a spectral blueshift.
\newpage

\newpage
\begin{center}
	\includegraphics[width=\textwidth]{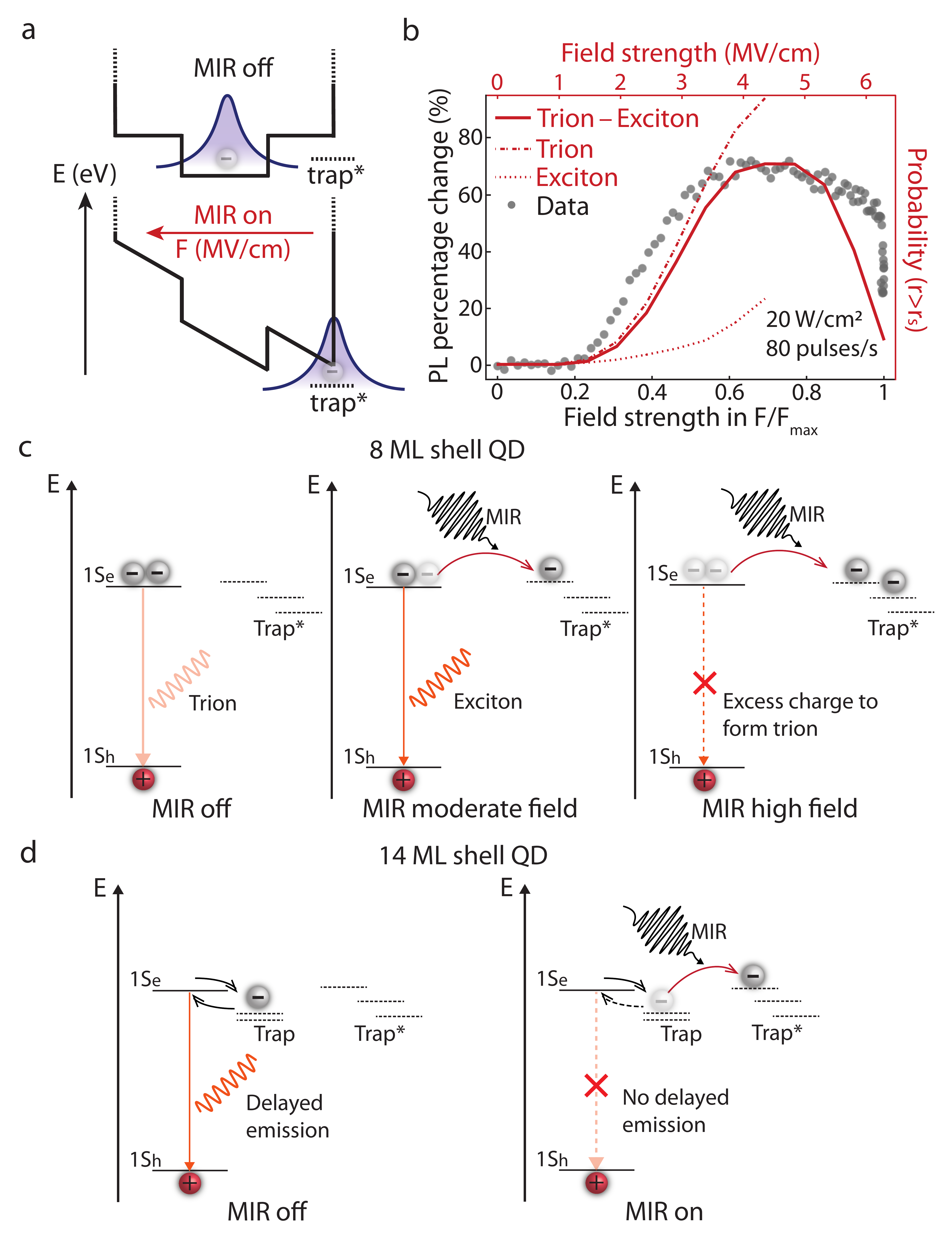}
	\label{fig:Fig5}
\end{center}
\item[Fig. 5] \textbf{Simulations and illustrations of the MIR-driven blinking control model.} \textbf{a,} Schematic depiction of the field-driven quantum tunneling: an external electric field perturbs the energy levels of trion states in the core, shell, and the whole QD, thereby reducing the potential barriers and assisting tunneling away from the recombination center. The exciton ionization and tunneling are modeled similarly by considering the exciton binding energy. \textbf{b,} Quantum tunneling simulation of the field dependence of MIR-induced PL enhancement. (The experimental data is reproduced from Fig. 2h). \textbf{c,} For 8 and 10 ML QDs, blinking OFF states in equilibrium originate from trion emission with low quantum yield. With moderate-field MIR excitations, the excess charge is removed to the trap states on the shell surface or outside environment (denoted as “trap*”), thereby restoring excitonic emission. At a high-field regime, MIR pulses not only remove the excess charge but also ionize the exciton, leaving an additional charge inside the dot. This extra charge leads to the formation of another trion and subsequent non-radiative Auger decay. \textbf{d,} In 14 ML QDs, electrons can be trapped in the shell denoted as “trap”, which can later thermally return to the recombination center and recombine radiatively, albeit slowly due to the giant shell. Under MIR irradiation, the trapped electron is pulled outside the dots (denoted as “trap*”) and cannot detrap back to the recombination center to form an exciton, leading to a quenched PL and leaving a charged QD behind.

\end{addendum}

\end{document}